\documentclass[prl,twocolumn,letterpaper,superscriptaddress]{revtex4}
\include{epsf}
\usepackage{graphicx}
\usepackage{pslatex}

\begin{document}
\title{New Limits on the Ultra-high Energy Cosmic Neutrino Flux from the ANITA Experiment}
\author{P.~W.~Gorham$^1$,
P.~Allison$^1$,
S.~W.~Barwick$^{2}$,
J.~J.~Beatty$^3$, 
D.~Z.~Besson$^4$,
W.~R.~Binns$^5$,
C.~Chen$^{12}$,
P.~Chen$^6$,
J.~M.~Clem$^7$,
A.~Connolly$^{11}$,
P.~F.~Dowkontt$^5$,
M.~A.~DuVernois$^9$, 
R.~C.~Field$^6$,
D.~Goldstein$^2$,
A.~Goodhue$^8$
C.~Hast$^6$,
C.~L.~Hebert$^1$,
S.~Hoover$^8$,
M.~H.~Israel$^5$,
J.~Kowalski,$^1$
J.~G.~Learned$^1$,
K.~M.~Liewer$^{10}$,
J.~T.~Link$^{1,13}$,
E.~Lusczek$^9$,
S.~Matsuno$^{1}$,
B.~C.~Mercurio$^3$,
C.~Miki$^{1}$,
P.~Mio\v{c}inovi\'c$^{1}$,
J.~Nam$^{2,12}$,
C.~J.~Naudet$^{10}$,
J.~Ng$^6$,
R.~J.~Nichol$^{11}$,
K.~Palladino$^3$,
K.~Reil$^6$,
A.~Romero-Wolf$^1$
M.~Rosen$^{1}$,
L.~Ruckman$^1$,
D.~Saltzberg$^8$,
D.~Seckel$^7$,
G.~S.~Varner$^{1}$,
D.~Walz$^6$,
Y.~Wang$^{12}$,
F.~Wu$^2$
}
\vspace{2mm}
\noindent
\affiliation{
Dept. of Physics and Astronomy, Univ. of Hawaii, Manoa, HI 96822. 
$^2$Dept. of Physics, Univ. of California, Irvine CA 92697. 
$^3$Dept. of Physics, Ohio State Univ., Columbus, OH 43210. 
$^4$Dept. of Physics and Astronomy, Univ. of Kansas, Lawrence, KS 66045. 
$^5$Dept. of Physics, Washington Univ. in St. Louis, MO 63130. 
$^6$Stanford Linear Accelerator Center, Menlo Park, CA, 94025. 
$^7$Dept. of Physics, Univ. of Delaware, Newark, DE 19716. 
$^8$Dept. of Physics and Astronomy, Univ. of California, Los Angeles, CA 90095.
$^9$School of Physics and Astronomy, Univ. of Minnesota, Minneapolis, MN 55455. 
$^{10}$Jet Propulsion Laboratory, Pasadena, CA 91109. 
$^{11}$Dept. of Physics, University College London, London, United Kingdom. 
$^{12}$Dept. of Physics, Grad. Inst. of Astrophys.,\& Leung Center for 
Cosmology and Particle Astrophysics, National Taiwan University, Taipei, Taiwan. 
$^{13}$Currently at NASA Goddard Space Flight Center, Greenbelt, MD, 20771.
}


\begin{abstract}
We report initial results of the 
first flight of the Antarctic Impulsive Transient Antenna (ANITA-1) 
2006-2007 Long Duration Balloon flight, which searched for evidence of 
a diffuse flux of cosmic neutrinos above energies of
$E_{\nu} \simeq 3 \times 10^{18}$~eV. ANITA-1 flew for 35 days looking for radio 
impulses due to the Askaryan effect in neutrino-induced
electromagnetic showers within the Antarctic ice sheets. We report here on
our initial analysis, which was performed as a blind search of the data. 
No neutrino candidates are seen, with no detected physics background.
We set model-independent limits based on this result.
Upper limits derived
from our analysis rule out the highest cosmogenic neutrino models.
In a background horizontal-polarization channel, we also
detect six events consistent with radio impulses from 
ultra-high energy extensive air showers.

\end{abstract}
\pacs{95.55.Vj, 98.70.Sa}
\narrowtext 

\maketitle

In all standard models for ultra-high energy cosmic ray (UHECR) propagation,
their range is ultimately limited by the opacity of the 
cosmic microwave background radiation. 
The UHECR energy above which this becomes significant is 
about $6 \times 10^{19}$~eV in the current epoch. This cuts off
their travel beyond distances of order 50~Mpc as first noted by
Greisen~\cite{Greisen}, and Zatseptin and Kuzmin~\cite{ZK} (GZK). As a result
of this absorption, the UHECR energy above this GZK cutoff 
is ultimately converted to photons, neutrinos, and lower energy hadrons.
The resulting neutrinos were first described by
Berezinsky and Zatsepin~(BZ)~\cite{BZ}. In standard UHECR source models
the BZ neutrino fluxes peak at energies about 2 orders of magnitude below the
GZK energy. Thus a ``guaranteed'' flux of neutrinos at energies of
$E_{\nu} = 10^{17-20}$~eV exists. Its detection is one of the clearest ways to
reveal the nature and 
cosmic distribution of the UHECR sources~\cite{Seckel05}, which is
one of the longest-standing problems in high energy astrophysics.

The ANITA-1 Long Duration Balloon experiment was designed specifically to search for this
cosmogenic BZ neutrino flux. ANITA-1 exploits the Askaryan effect, in which
strong coherent radio emission arises from electromagnetic showers in
any dielectric medium~\cite{Ask62}.
The effect was first observed in 2000~\cite{Sal01}, and
has now been clearly
confirmed and characterized for ice as the medium, as part of the pre-flight calibration
of the ANITA-1 payload~\cite{slac07}. A prior flight of a prototype payload
called ANITA-lite in 2003-2004 led to validation of the technique and initial
neutrino flux limits that ruled out several UHE neutrino models~\cite{anitalite}.

In a previous paper~\cite{ANITA-1}, we describe
in detail the ANITA-1 instrument, payload, and flight system. Reference ~\cite{ANITA-1}
also includes
details of the instrument performance during the flight, estimates of the overall
sensitivity of the instrument to neutrino fluxes, and discussions of possible
backgrounds. Because of the complexity of the flight system and methodology, 
we refer the reader to ref.~\cite{ANITA-1} for more detail when necessary.

\begin{figure}[ht!]
\begin{center}
\includegraphics[width=3.35in]{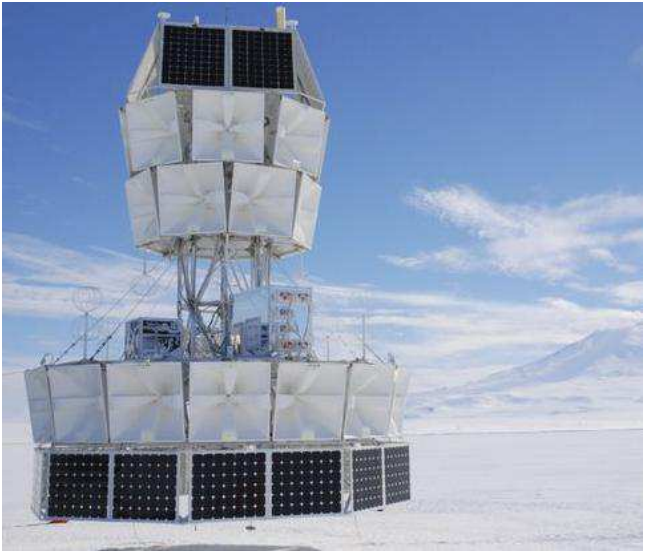}
\caption{View of the ANITA-1 payload in launch configuration, with
photovoltaics at the top and bottom, and antenna clusters between. The
side of each square antenna mouth is about 0.9~m, and the payload stands
about 8~m tall.}
\label{payload}
\end{center}
\end{figure}

The ANITA-1 payload (Fig.~\ref{payload})
launched from Williams Field, Antarctica
 near McMurdo station, on December 15, 2006, and executed
more than three circuits of the continent. The payload landed on the Antarctic
plateau about 300~miles from Amundsen-Scott (South Pole) Station, after 35 days aloft. Anomalous
stratospheric conditions led to a misalignment of the polar
vortex for the 2006-2007 season, and as a result the ANITA-1 trajectory spent
an unusually large fraction of the time over West Antarctica where the ice sheet
is smaller and shallower. In addition, the payload field-of-view to the horizon
(at a distance of about 650~km at typical altitudes of 35-37~km above mean sea level,
or 33-35~km above the ice surface),
often included the two largest occupied stations in Antarctica, McMurdo and
Amundsen-Scott, and thus was subject to higher-than-expected levels of anthropogenic
electromagnetic interference (EMI). 
Despite these effects, the payload 
accumulated a net exposure livetime of 17.3 days with a mean ice depth in the
field of view of 1.2~km, comparable to the attenuation length of the ice at sub-GHz radio
frequencies~\cite{icepaper}. ANITA-1 was
thus able to synoptically view a volume of ice of $\sim 1.6$~M~km$^3$. 
Our volumetric acceptance to a diffuse
neutrino flux, accounting for the small solid angle of acceptance for any given volume element, is
several hundred km$^3$ water-equivalent steradians at $E_{\nu} \simeq 10^{19}$~eV~\cite{ANITA-1}.

ANITA-1's antennas are 32 dual-linear-polarization,
quad-ridged horns, each
with a field of view which averages about $50^{\circ}$ angular diameter over their 200-1200~MHz
working bandwidth. The antennas are arranged in upper and lower rings,
each with 16 antennas at azimuthal intervals of $22.5^{\circ}$. All antennas
point at $10^{\circ}$ below the horizontal, to maximize sensitivity to
the largest portion of the volume near the horizon at $6^{\circ}$ below the
horizontal. The combined view of all antennas covers the entire lower hemisphere
down to nadir angles of about $55^{\circ}$, comprising 99.4\% of the area
within the horizon.  Radio impulses that exceed the ambient thermal noise by about
$5\sigma$ in at least 
four antennas in coincident upper- and lower-ring pairs produce a trigger~\cite{ANITA-1},
and the entire antenna set of waveforms are then digitized and stored for
later analysis. Thermal noise fluctuations produce
random triggers at a rate of about 4-5~Hz, yielding a continuous monitor of 
instrument health. These events are incoherent in phase and produce
a completely negligible background to actual coherent radio impulses.



The event analysis is conceptually simple, but requires detailed calibration
of the instrument to achieve good precision. In the results reported here,
we accepted only events having at least six adjacent
antennas with detectable signals. The six antenna
signals are analyzed using a method of pulse-phase interferometry to
determine the best arrival direction of the radio impulse plane wave,
and this direction and its associated uncertainty is then mapped onto
the Antarctic ice surface by reference to 
onboard payload navigation instruments, with an angular precision
of $0.2^{\circ} \times 0.8^{\circ}$ in elevation and azimuth~\cite{ANITA-1}.

\begin{table}[htb!]
\caption{\label{event-table}Event totals vs. analysis cuts and estimated signal 
efficiencies for unblinded ANITA-1 data set.
}
\begin{center}
\begin{tabular}{lrrrc}
\hline
Cut requirement~~~~~~~~~~passed: & total & Hpol & Vpol & Efficiency\\ \hline
(0) Hardware-Triggered& $\sim 8.2$M & ... & ... & ... \\
(1) Upcoming plane wave & 32308 & 15997 & 16311 & 0.93 \\
(2) Impulsive broadband~~ & 19695 & 10095 & 9600 & 0.98\\
(3) Isolated from other events & 9 & 8 & 1 & 0.94 \\ 
(4) Isolated from camps & 6 & 6 & 0 & 0.96 \\
(5) Vpol dominant & 0 &  0 & 0  & 0.99\\ 
\hline
\end{tabular}
\end{center}
\end{table}

\begin{figure*}[ht!]
\begin{center}
\centerline{\includegraphics[width=3.3in]{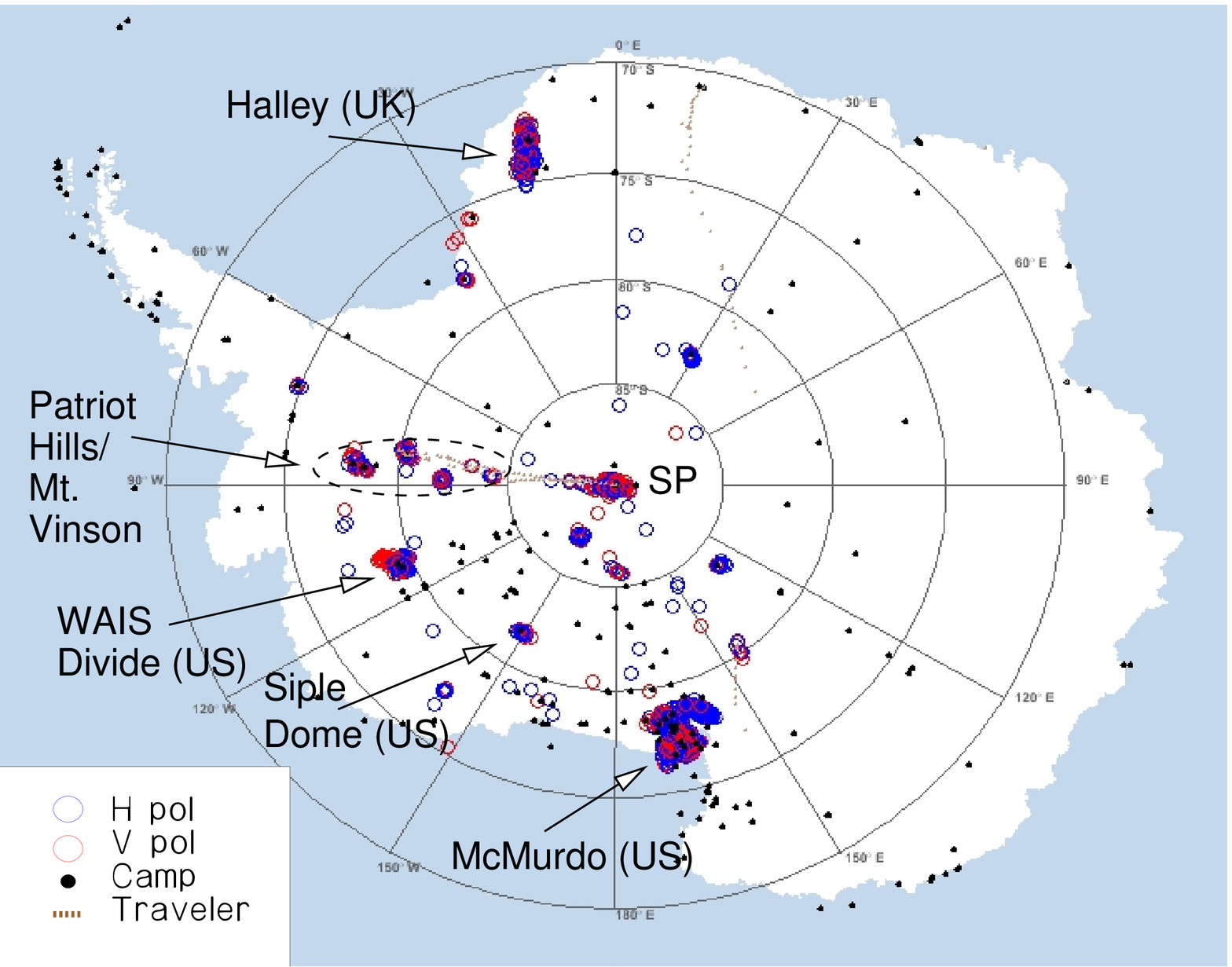}\includegraphics[width=3.3in]{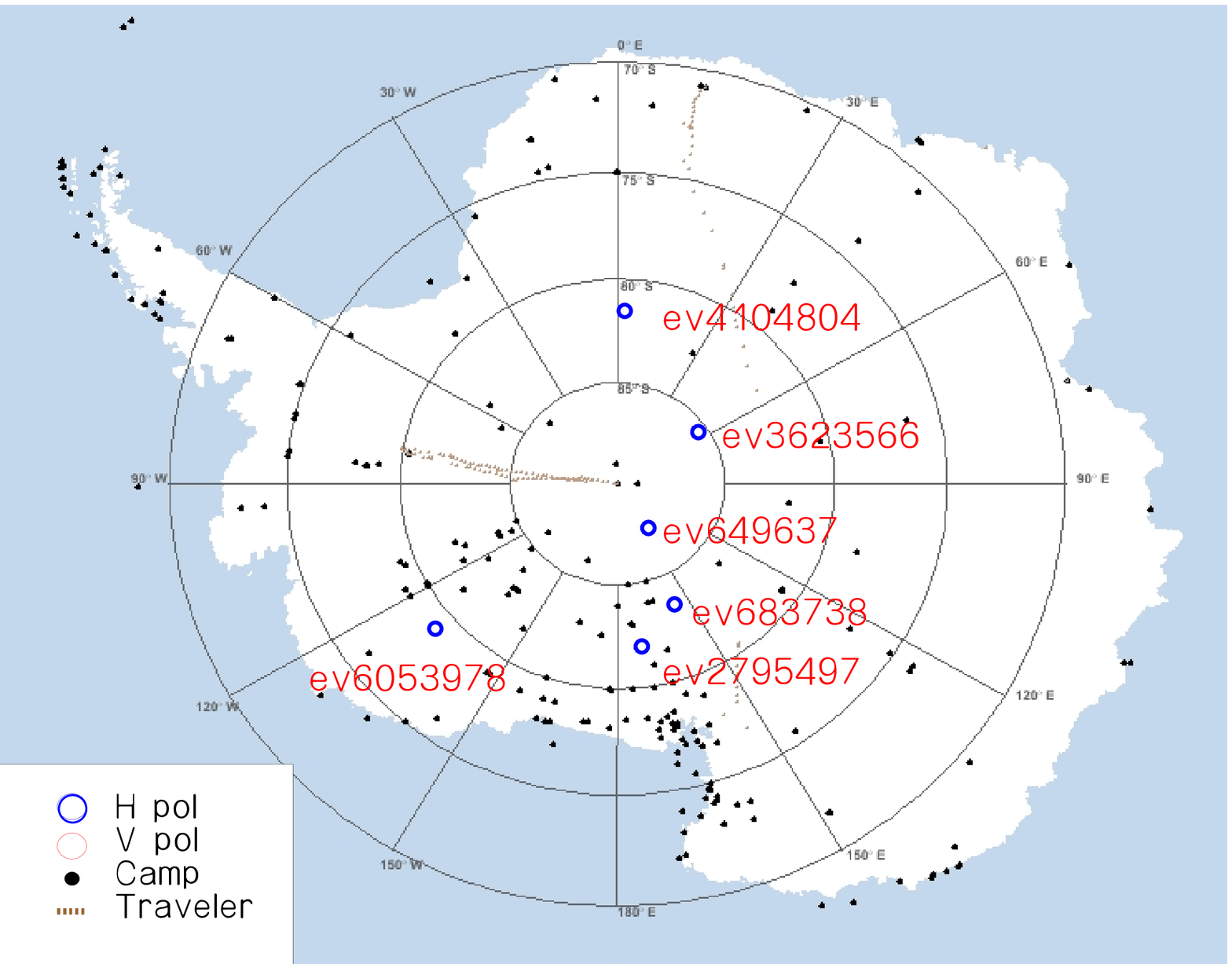}} 
\caption{Left: Plot of all reconstructed events, in both horizontal and vertical polarization;
major Antarctic stations are indicated on the map.
Right: events remaining after cuts to remove anthropogenic interference. 
6 events remain in the horizontally
polarized group, but these are non-candidates for neutrino events, as discussed in the text.}
\label{recons}
\end{center}
\end{figure*}

To minimize bias, the analysis cuts were
optimized on a 10\% randomized sample of the entire data set, and the remaining
90\% was blinded from the analysts until the cuts were fixed. 
The cuts proceed as follows:
{\bf (1)} Events that do not reconstruct to a coherent plane wave in arrival direction
are rejected as random thermal noise;
events that reconstruct from directions above the horizon are also rejected.
{\bf (2)} Events that reconstruct but have non-impulsive waveforms from relatively narrow-band
sources ($\leq 100$~MHz) are rejected.
{\bf (3)} Events that cluster with one another in source location 
to within reconstruction errors projected onto
the ice, or 50 km radius, whichever is greater, are rejected. 
True source candidates must be single, isolated events. Note
that this cut, and the ``camp cut'' that follows it, are largely but not completely redundant.
{\bf (4)} Events that coincide in source location with any known active or inactive
station, camp, aircraft flight path, or expedition traverse path, to within reconstruction 
angular errors projected onto
the ice, or 50 km radius, whichever is greater, are rejected as being associated with
anthropogenic activity. Even inactive camps or those long-abandoned are considered
a risk, since left-over equipment might serve as a site for charge accumulation and
associated electromagnetic discharges which could be mistaken for signals.
{\bf (5)} Events whose radio waveforms are not predominantly vertically polarized (Vpol) 
are rejected because,
from considerations of the Askaryan impulse generation process, and the Fresnel transmission
through the ice surface, they cannot originate from a 
particle shower within the ice sheet. Conversely, strongly horizontally
polarized (Hpol) events are likely to originate from above the ice from similar considerations.

Table~\ref{event-table} shows the results of the total event sample after unblinding,
including the signal efficiency for each cut separately. The
10\% initial sample is included in the totals. Note that the isolation cut (3)
is the single most stringent criterion in rejecting impulsive events, and this shows that the
vast majority of triggers are not single, isolated events.  Signal efficiency in each
case was tested with a simulated event sample injected randomly into the data
stream, and the final energy-averaged efficiency of all cuts is estimated to be 81\%.

In Figure~\ref{recons} we show the before-and-after maps 
of reconstructed ANITA-1 events superposed on the
Antarctic continent. The strong correlation to a small number of stations is evident.
The 6 surviving Hpol
events are by contrast widely distributed across the continent,
with no known camps or bases, either current or former, anywhere in their locale.
We have investigated the possibility of impulsive signals from earth-orbiting satellites
seen in reflection off the ice surface as a source for these events.
This hypothesis is ruled out because the waveforms for these events do not show any evidence of
differential group delay from ionospheric dispersion
which is several ns per MHz in the 200-400~MHz frequency range where
these events have most of their 
spectral power. In fact the signals are
all of durations less than 10~ns.  We know of no other anthropogenic sources
for these events.


With regard to possible physics sources, our simulations of the high-frequency tail 
of impulsive geo-synchrotron radio emission~\cite{Huege_Falcke05,Falcke,Suprun}
from  ultra-high energy cosmic ray extensive air showers
(EAS) suggest that these signals may be EAS events 
seen in reflection off the ice surface~\cite{ANITA-1}. 
Such events are expected to be predominantly
Hpol because of the strong Fresnel reflectivity in the 
region near Brewster's angle, and the overall
initial preference for Hpol because of the more vertical polar magnetic fields. 
Our simulations
predict a handful of such events for the flight, all of which arise from UHECR EAS with energies
above $10^{19}$~eV; however, the uncertainties are large~\cite{ANITA-1}.
While these events do not constitute a background for our neutrino search because
of their incorrect polarization, they are a potentially interesting signal in their own right. 
Further analysis, including a search for similar events from above the horizon,
is in progress.

\begin{figure}[ht!]
\begin{center}
\includegraphics[width=3.2in]{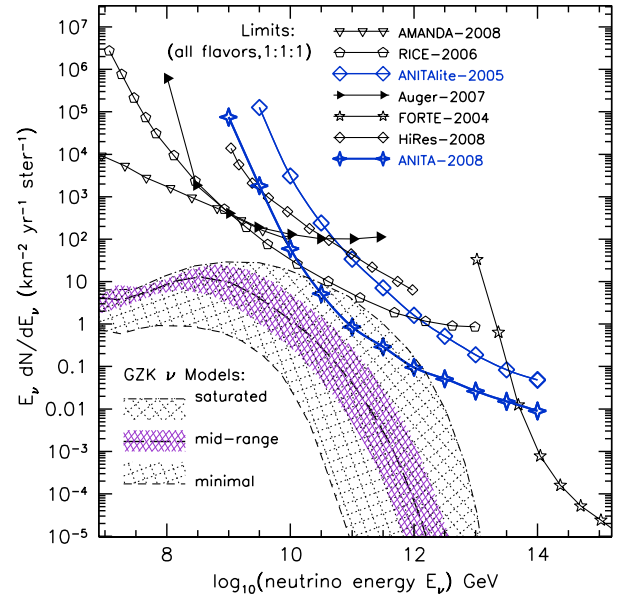} \\
\caption{ANITA-1 limits based on no surviving candidates for 18 days of livetime. Other
limits are from AMANDA~\cite{AMANDA08}, RICE~\cite{RICE06}, ANITA-lite~\cite{anitalite},
Auger~\cite{Auger07}, HiRes~\cite{Hires08}, FORTE~\cite{FORTE04}. The BZ (GZK) neutrino
model range is determined by a variety of 
models~\cite{PJ96,Engel01,Kal02,Kal02a,Aramo05,Ave05,Barger06}.}
\label{lim08}
\end{center}
\end{figure}

\begin{table}[hbt!]
\caption{Expected numbers of events $N_{\nu}$ from several UHE neutrino models, and 
confidence level $CL=100(1-exp(-N_{\nu}))$ for exclusion by ANITA-1 observations.
\label{table2}}
\vspace{3mm}
 \begin{footnotesize}
  \begin{tabular}{lcr}
\hline \hline
{ {\bf Model \& references}}   &  predicted {$N_{\nu}$}~~      &  ~~{\bf CL,\%} \\ \hline
{\it Baseline BZ models} &  &  \\
~~~~~~Protheroe \& Johnson 1996~\cite{PJ96} & 0.22  & 19.7 \\
~~~~~~Engel, Seckel, Stanev 2001~\cite{Engel01} &  0.12 & 11.3 \\
~~~~~~Barger, Huber, \& Marfatia 2006~\cite{Barger06} & 0.38 & 31.6\\
{\it Strong source evolution BZ models}&  &  \\
~~~~~~Engel, Seckel, Stanev 2001~\cite{Engel01} & 0.39 & 32.3\\
~~~~~~Kalashev {\it et al.} 2002~\cite{Kal02} &  1.03 & 64.3 \\
~~~~~~Aramo {\it et al.} 2005~\cite{Aramo05} & 1.04 & 64.6\\
~~~~~~Barger, Huber, \& Marfatia 2006~\cite{Barger06} & 0.89 & 58.9\\
~~~~~~Yuksel \& Kistler 2007~\cite{Yuksel07} & 0.56 & 42.9\\
{\it BZ Models that saturate all bounds}: & & \\
~~~~~~Kalashev {\it et al.} 2002~\cite{Kal02} & 10.1 & $>99.99$\\
~~~~~~Aramo {\it et al.} 2005~\cite{Aramo05} & 8.50 & $>99.98$\\
{\it Waxman-Bahcall fluxes}: & & \\
~~~~~~Waxman, Bahcall 1999, evolved sources~\cite{WB}~~ & 0.76 & 53.2 \\
~~~~~~Waxman, Bahcall 1999, standard~\cite{WB} & 0.27 & 23.7\\ \hline
  \end{tabular}
 \end{footnotesize}
\end{table}

Based on the approach described in Refs.~\cite{Anch02,FORTE04}, the
resulting model-independent 90\%~CL limit on neutrino fluxes with Standard 
Model cross-sections~\cite{Gan00} is shown in Fig.~\ref{lim08}.
Here we have included the net livetime and 81\% analysis efficiency.
Exclusion of the volume of ice near all camps and events 
reduces the net effective volume by a
few percent. We estimate that experimental
systematics such as variations in ice properties and 
calibration uncertainties in
the absolute radio signal strength lead to uncertainties 
of order a factor of two on the limit.
These model-independent limits are calibrated such that a model spectrum
that matched the limit over one decade of energy would yield approximately
2.3 events; this choice is appropriate to smoothly varying models.
The limits are an average over all three neutrino flavors, as ANITA-1 had
roughly equal sensitivity to $\nu_e,~\nu_{\mu},~\nu_{\tau}$, and
the flavors should be equally mixed to first order via oscillations for all
BZ neutrino models.
We plot only an approximate set of bands for the BZ neutrino models, 
which are too numerous to 
individually plot here.

In Table~\ref{table2} we give the total number of events expected from 
selected individual ultra-high energy neutrino models which 
are representative of the range of BZ neutrino expectations; we also
include the case of a model which saturates
the canonical Waxman-Bahcall flux bounds for both evolved and standard
UHECR sources~\cite{WB}. ANITA-1 strongly limits the highest
BZ neutrino models, which require extremely high-energy
cutoffs in the parent cosmic-ray sources spectral energy distribution.
Our limits thus suggest that UHECR source spectra extending to
$10^{23}$~eV are disfavored. ANITA-1 sensitivity approaches
a class of models here denoted as ``strong-source evolution'' models,
which assume that the UHECR source evolution follows the cosmic
evolution of more energetic sources, for example 
gamma-ray burst host galaxies~\cite{Yuksel07}; 
these mid-range models are constrained at about the 60\% CL
but none are ruled out yet.
The ANITA-1 90\% CL integral flux limit on a pure $E^{-2}$ spectrum
for the energy range $10^{18.5}$~eV $\leq E_{\nu} \leq 10^{23.5}$~eV is 
$E_{\nu}^2 F \leq 2 \times 10^{-7}$ GeV cm$^{-2}$ s$^{-1}$ sr$^{-1}$.

In summary, we have set the strongest bounds to date on the ultra-high
energy neutrino flux at energies above $3 \times 10^{18}$~eV, using
the radio Cherenkov method via synoptic observations of the Antarctic
ice sheets from stratospheric altitudes. Our methodology appears
to have no observed physics backgrounds, but may have detected events
due to cosmic-ray extensive air showers that are easily
separated from the neutrino events we seek. 

This work has been supported by the National Aeronautics and Space
Administration, the National Science Foundation Office of Polar Programs, 
the Department of Energy Office of Science High Energy
Physics Division, and the UK Science
and Technology Facilities Council. Special thanks to the staff of 
the Columbia Scientific Balloon Facility.

\end{document}